\documentclass[usenatbib]{mn2e}
\usepackage{mn2e_journal_style}
\usepackage{graphicx}
\usepackage{amssymb}
\newlength{\figurewidth}
\newcommand{\revised}[1]{#1}
\title[Older stars in the ONC]%
{Captured older stars as the reason for apparently prolonged star formation
in young star clusters}
\author[J.~Pflamm-Altenburg, P.~Kroupa]
{Jan~Pflamm-Altenburg$^1$$^2$\thanks{email: jpflamm@astro.uni-bonn.de,
    pavel@astro.uni-bonn.de}
  and Pavel~Kroupa$^1$$^2$\footnotemark[1]\\
  $^1$ Argelander-Institut f\"ur Astronomie 
  \thanks{Founded by 
    merging of the Institut f\"ur Astrophysik
    und Extraterrestrische Forschung, the Sternwarte, and the
    Radioastronomisches Institut der Universit\"at Bonn.}
  Universit\"at Bonn
  , Auf dem H\"ugel 71, D-53121 Bonn\\
  $^2$ Rhine Stellar Dynamics Network (RSDN)
}
\date{Accepted 2006 November 15}
\begin{document}
\maketitle
\begin{abstract}
  The existence of older stars within a young star cluster
  can be interpreted to imply that star formation occurs on 
  time scales longer than a free-fall time of a pre-cluster cloud core.
  Here the idea is explored that these older stars
  are not related to the star formation process forming the 
  young star cluster but rather that the orbits of older field stars 
  are focused
  by the  collapsing pre-cluster cloud core. Two effects appear:  
  The focussing  of stellar orbits leads to an enhancement 
  of the density of field stars in the vicinity of the
  centre of the young star cluster. 
  And due to the time-dependent potential of the forming cluster
  some of these stars can get bound gravitationally to the cluster. 
  These stars
  exhibit similar
  kinematical properties
  as the newly formed stars and can not be distinguished from them
  on the basis of radial-velocity or proper-motion surveys. Such 
  contaminations may lead to a wrong apparent star-formation history
  of a young cluster. In the case of the ONC the theoretical
  number of gravitationally
  bound older low-mass field stars agrees with the number of observed older
  low-mass stars.
  
\end{abstract}
\section{Introduction}
  \citet{palla2005a} determined the ages of 84 low-mass
  ($m~\approx~0.4\--1.0\;\mathrm{M}_\odot$) stars
  in the Orion Nebula cluster (ONC) from isochrones and
  lithium depletion. Four of these stars have ages
  between 10 and 18~Myr, whereas the bulk of all
  stars have ages below 3~Myr. They conclude that stars in the
  ONC formed moderatly over
  a long time period exceeding 10~Myr ending in a sharp 
  peak of star formation.   
  \revised{This age spread of stars in the ONC has already been recognised by
    \cite{isobe1982a}, who determined ages from 10$^4$~yr up to 30~Myr and
    that low-mass stars in the Orion Nebula region have the same ages as 
    the oldest stars in the Orion association Ia.}
  
  This contradicts recent results that the process of star formation is 
  rapid.
  In the case of the Taurus star forming region \citet{hartmann2003a}
  has shown that after correcting 
  the sample of stars for possible foreground contamination, the age spread 
  narrows.
  Also, observations  suggest that star formation occurs in only one or 
  two crossing times \citep{elmegreen2000a} which results in age spreads
  much smaller than 10~Myr.

  Contrary to this, \citet*{tan2006a} present observational and
  theoretical arguments that rich star clusters requires at least several
  dynamical time-scales to form and they are quasi-equilibrium structures
  during their assembly. For the ONC they concluded that it has formed
  over $\gtrsim$~3--4 dynamical times.
  
  Indeed, the measurements by \citet{palla2005a} are excellent, comparing
  ages of stars in the ONC derived with different methods. 
  But the existence of older stars in the ONC must not stringently imply
  that star formation is an extended process. Given the velocity dispersion
  of the low-mass stars in the ONC is approximately 2.5~km~s$^{-1}$ 
  \citep{hillenbrand1998a} a star not originating from the ONC
  can come from a place up to 25~pc away if it is 10~Myr older than
  the cluster member stars.

  In the case of $\omega$ Cen ($2.5\cdot 10^{6}$~M$_\odot$)
  \citet*{fellhauer2006a} have
  shown that a massive stellar super-cluster may trap 
  older galactic field stars during its formation process that are 
  later detectable in the cluster as an apparent population of stars 
  with a very different age and metallicity.  Up to about 40\% 
  of its initial mass can be additionally gained from trapped disc stars,
  while certain conditions may even lead to such a massive cluster 
  capturing a multiple of its own mass.

  Here we show that a collapsing pre-ONC cloud core leads to an
  enhancement of fore- and background stars and to the capture of some
  older field stars so that they are gravitationally bound by the new 
  star cluster. Given realistic initial conditions the order 
  of magnitude of the number of captured stars 
  agrees with the number of older low-mass stars in the ONC. 
  These captured older stars have similar radial velocities and 
  proper motions as the newly formed stars in the ONC and can 
  therefore not be distinguished 
  kinematically from the new stars, which will lead to a much wider derived
  age spread.

\section{Model}
The entire model consists of a collapsing pre-cluster cloud 
core embedded in
a uniform field population of older stars, which --- in the  case of the ONC
--- can be a slowly expanding
association.

The collapsing pre-cluster cloud core is described by a 
time-dependent Plummer potential
\begin{equation}
\Phi_{\mathrm{cl}}(t) = -M_{\mathrm{cl}}\;G\;
\left(r^2+b^2_{\mathrm{cl}}(t)\right)^{-\frac{1}{2}},
\end{equation}
where $M_\mathrm{cl}$ is the constant total mass 
of the collapsing cloud. The Plummer parameter, $b_\mathrm{cl}(t)$, 
is a function of time to describe the growing potential
starting with an infinite value and ending 
with a finite final value within a finite time to account for 
the ongoing concentration of the pre-cluster cloud core forming a new star
cluster.

The background stars are set up in a sphere centred on
the origin of the Plummer potential.
The positions of the stars are uniformly distributed and
their initially isotropic
velocities have a  Gaussian distribution. Choosing a field-sphere
is uncritical as long as its radius is large enough.

In this model the stars do not interact gravitationally. They 
move as test particles in an external potential. 
The stellar masses can be eliminated in the equation of
motion and the result does not depend on the masses of the stars.

The orbit of the field stars are integrated using a 
standard Hermite scheme
\citep{makino1991a,hut1995a,aarseth2003a}.
The total energy is used to control the integration. Due to the 
time-dependent cloud potential the total energy is not conserved.
Therefore the path integral, 
\begin{equation}
W = \int \bmath{F}_{\mathrm{cl}}\bmath{\cdot}\mathrm{d}\bmath{r}\;\;\;,
\end{equation} 
of the stars in the force field of
the cloud is computed. Given the energies,
\begin{equation} 
E_0 = U_{0} + T_0\;\;\; 
\mathrm{and} \;\;\;
E_t = U_{t} + T_t\;\;\;,
\end{equation} 
where $E_0$, $U_0$ and $T_0$ are the total, potential and kinetic
energies at the initial time $t_0$ and $E_t$, $U_t$ and $T_t$
are the total, potential and kinetic
energies at an arbitrary time point $t$,
we control the calculations by the total energy error
\begin{equation}
\epsilon = \frac{E_0 - \left(E_t - W\right)}{E_0}.
\end{equation}
In all simulations the total relative energy error is less 
than 10$^{-11}$ per particle.
\section{Initial Conditions} 

To simulate the capture of stars by the above model, initial conditions
for the underlying stellar field population and the collapsing cloud must be
specified. 
In the case of the ONC the stellar background population
is given by a postulated surrounding association.  
\subsection{Collapsing cloud}
According to \citet{hillenbrand1998a}
the virialized total mass of the ONC is determined to be about 
4500~M$_\odot$ while only one half is visible in stellar material.
If virialisation of the very young ONC is assumed 
before gas expulsion has started 
then the cloud mass, $M_{\mathrm{cl}}$, can be set to
4500~M$_\odot$.
  
The cloud collapse begins at time $t_b$ and ends at $t_e$ after the 
collapse time, $\tau_\mathrm{cl}$, has elapsed.
For a constant cloud mass the increasing compactness
of the collapsing cloud is described by a Plummer parameter
of $+\infty$ before the collapse starts and has the constant value $b_\mathrm{0}$
when the collapse finishes. Between these points the Plummer parameter
is interpolated simply by 
\begin{equation}
b_{\mathrm{cl}}(t) = \left\{
\begin{array}{l@{\;\;\;;\;\;\;}l}
+\infty & t < t_b\;\;\;,\\
b_0\;\frac{(t_e-t_b)}{t-t_b} & t_b\le t \le t_e\;\;\;,\\
b_0 & t> t_e\;\;\;,\\
\end{array}\right.
\end{equation}
where $t_e-t_b$ is the collapse time  $\tau_{\mathrm{c}}$.
\citet{hillenbrand1998a} specify the core radius 
(projected half-density radius)
of the current ONC to lie
between 0.16 and 0.21~pc. The actual core radius $r_\mathrm{c}$ 
and the Plummer radius $b_0$ are related by 
\begin{equation}
r_{\mathrm{c}} = \left(\sqrt{2} -1 \right)^{\frac{1}{2}}\;b_0 \approx
0.64\;b_0
\;\;\;,
\end{equation}
which implies a current Plummer radius for the ONC between 0.25 and 0.33~pc.
Here we choose $b_0=0.30$~pc for the ONC as a mean value.

The collapse time, $\tau_{\mathrm{c}}$, of the pre-cluster cloud core is 
estimated by the free-fall time scale \citep{elmegreen2000a}
\begin{equation}
t_{\mathrm{ff}} \approx \sqrt{\frac{R³}{G\;M_{\mathrm{cl}}}}\;\;\;.
\end{equation}
If the extension of the cloud at the onset of collapse was 1~pc, the
corresponding free-fall time is computed to be about 0.22~Myr, in the case
of 5~pc as the initial radius the free-fall time-scale is 2.50~Myr and 7.07~Myr
for a 10~pc radius.
By measuring
the offset between HII- and CO-arms in spiral galaxies \citet{egusa2004a}
determined the time for star clusters to ''hatch'' from their natal 
cluster to be about 5~Myr. Also \citet*{weidner2004b} conclude from a
comparison of star formation rates and maximum cluster masses 
in a large ensemble of galaxies that
pre-cluster cloud cores have radii of about 5~pc if they form in a free
fall period. Because the possible collapse time-scale can vary over 2~decades
the collapse time $\tau_\mathrm{c}$ is taken here to vary 
between 0.1 up and 10.0~Myr.

\subsection{Underlying field population}
\label{underlying_field_population}
The task to make a specific setup for the underlying field population
at the onset of the collapse is far more difficult than for 
the collapsing cloud.
\revised{The conditions of the stellar density before collapse have to be
estimated.  For example, the stellar local mass density of 
the solar neighbourhood 
is about  0.1~M$_\odot$~pc$^{-3}$ 
\citep*{bahcall1992a,bienayme1987a,kuijken1989a,kroupa1993a}.
In the compiled radial morphology around $\Theta^{1}$C 
(\citealt{hillenbrand1998a,herbig1986a}) the outermost 
population (the Orion Ic association) has an extent of more than 25 pc. 
The embedding cloud Orion A also contains a large number of small groups
and a significant distributed population \citep{megeath2005a,strom1993a}. 
The  ONC is thus part of a region with low- and high mass
star formation in the recent past and at present. 
So it can be assumed that the stellar density must have been higher
at the onset of the pre-cluster cloud collapse than in the solar vicinity.
We choose the background as a sphere with a radius of 12.5~pc, corresponding
to the extent of the embedding association. 2410 stars distributed in this
sphere would give a mean number density of 0.3~stars~pc$^{-3}$ or 
a mass density of 0.1~M$_\odot$~pc$^{-3}$ equal to the mean stellar 
mass density of the solar vicinity.  
Here, the number of background particles is rather taken to be
20000 on two grounds:
the resulting density, 2.44~stars~pc$^{-3}$ or 0.83~M$_\odot$~pc$^{-3}$,
is slightly higher than the density in the solar vicinity and 
this high number of particles guaranties useful statistical results.
These 20000 stars could have been formed e.g. in two ONC-type star clusters.

In fact, the total number of stars within the setup
sphere and its radius are not the primary parameters
affecting the results, but rather the resulting number density and
the initial velocity dispersion. As these stars act as test particles 
the results can be scaled linearly with the initial uniform density.

For the brightest members of
the ONC \citet{vanaltena1988a} found a one-dimensional velocity dispersion
of 1.49~km~s$^{-1}$, while \citet{jones1988a} specify the velocity
dispersion  to be slightly larger with 2.34~km~s$^{-1}$. 
In \citet{hillenbrand1998a} a value of 2.81 for stellar masses between
0.1 $<$ $m/$M$_\odot$ $<$ 0.3 and 2.24~km~s$^{-1}$ between
1 $<$ $m/$M$_\odot$ $<$ 3 is reported. If the underlying 
stellar background population had a similar progenitor the velocity
dispersion can be assumed to be of the same order, i.e. 
1$\sim$3~km~s$^{-1}$. Thus we vary the one dimensional 
initial velocity dispersion of the background sphere from 
0.5 (0.87) up to 3 (5.20)~km~s$^{-1}$ 
(three dimensional dispersion in parentheses).

Summarising, the choice of the initial background population is a follows:
20000 particles are uniformly distributed over a sphere with a
radius of 12.5~pc, giving a stellar density of 2.4~stars~pc$^{-3}$.
The particles have a Maxwellian velocity distribution 
and a random direction. The velocity dispersion, $\sigma$,
of the different models varies from 0.5~km~s$^{-1}$ to
3.0~km~s$^{-1}$. The background sphere and the collapsing cloud 
have the same velocity centroid. 
}

\section{Stellar capture}
\subsection{Number of expected captured stars and the IMF}
\citet{palla2005a} selected a sample of 84 stars in the range
$\approx$0.4--1.0~M$_\odot$ and with isochronal ages greater than 
$\sim$~1~Myr  out of the ONC-survey made by \citet{hillenbrand1997a}.
This ONC-survey covers 3500 stars within 2.5~pc of the 
central Trapezium. The low-mass stars of the sample
have a membership probability greater than 90 per cent. 
6~stars (7.1 per cent)  of the sample of 84~stars
have isochronal ages $\gtrsim$~10~Myr, whereas four of them show a significant
lithium depletion. The ages derived from the amount of lithium depletion
confirm the ages derived from isochronal lines.   

To estimate the expected total number 
of older stars in the ONC this sample must be extrapolated to the entire
ONC. The actual total mass of the ONC is given by \citet{hillenbrand1998a}
to be about 1800~M$_\odot$. Using the universal 
or standard/canonical  IMF 
\citep{kroupa2001a,weidner2006a,pflamm-altenburg2006a} 
and the ''WK-normalisation'' method
the ONC 
should have formed 694 low-mass stars in the mass regime 0.4--1.0~M$_\odot$.
Note that the ''WK-normalisation'' refers to the maximum mass of the star
being determined by the cluster mass  \citep{weidner2004a,weidner2006a}. 
This determines the normalisation 
constant of the IMF. 
\revised{Given 6 older stars out of a sample of 84 stars, then 78 stars
of this sample should have formed in the ONC. Thus, 
$6/78 \times 694 \approx 53$
older stars are expected among 694 newly formed stars after linear
extrapolation.}
 
\subsection{Calculated  number of captured stars}
After the collapse time $\tau_\mathrm{c}$, when the collapse has stopped,
then a star is identified to be captured by the collapsed cloud 
if the star is gravitationally bound and lies within a 2.5 pc~radius
of the centre of the potential,
according to the extend of the ONC-survey by \citet{hillenbrand1997a}:
after the collapse stops the distance of the star to the 
origin of the Plummer sphere is less than 2.5~pc and 
the total energy of the stars is less than required to get 
farther away than 2.5~pc from the centre of the potential.
\revised{This means that a captured star is gravitationally bound to the
new cluster. As the virial mass of the current ONC is about 4500~M$_\odot$
but only one half is visible in stellar material \citep{hillenbrand1998a},
the ONC is super-virial (kinematically too hot). 
This can be solved if the ONC is expanding after gas expulsion 
\citep*{kroupa1999a,kroupa2001b}. 
The whole cluster 
is expected to have been virialized between
the stop of the pre-cluster cloud core collapse and the start 
of gas expulsion.  After gas
expulsion the older captured stars follow the dynamical evolution of the young
star cluster. So it is justified to identify a captured star by the 
criterion that it be gravitationally bound to the collapsed cloud when 
the collapse of the pre-cluster cloud core stops.
}

\begin{figure}
  \includegraphics[width=\figurewidth]{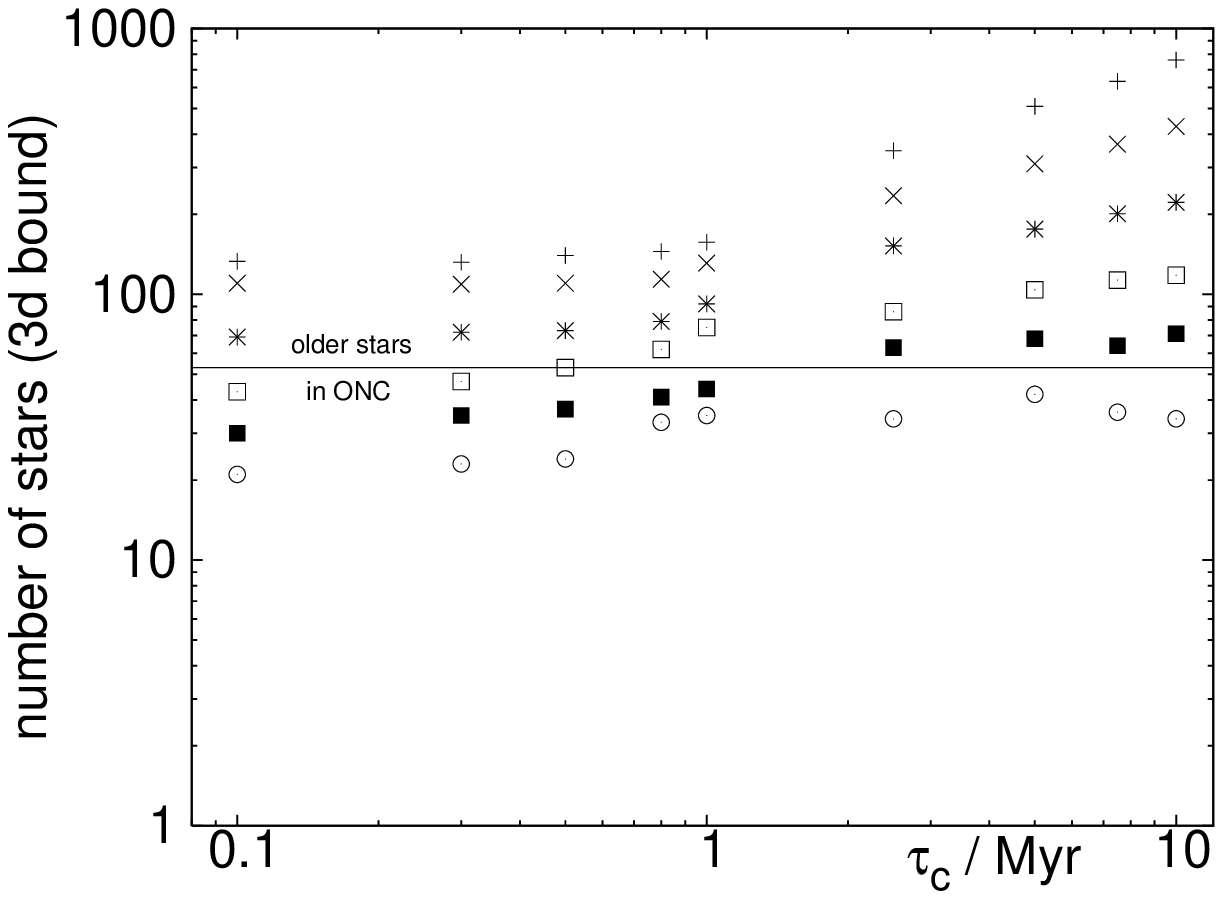}
  \includegraphics[width=\figurewidth]{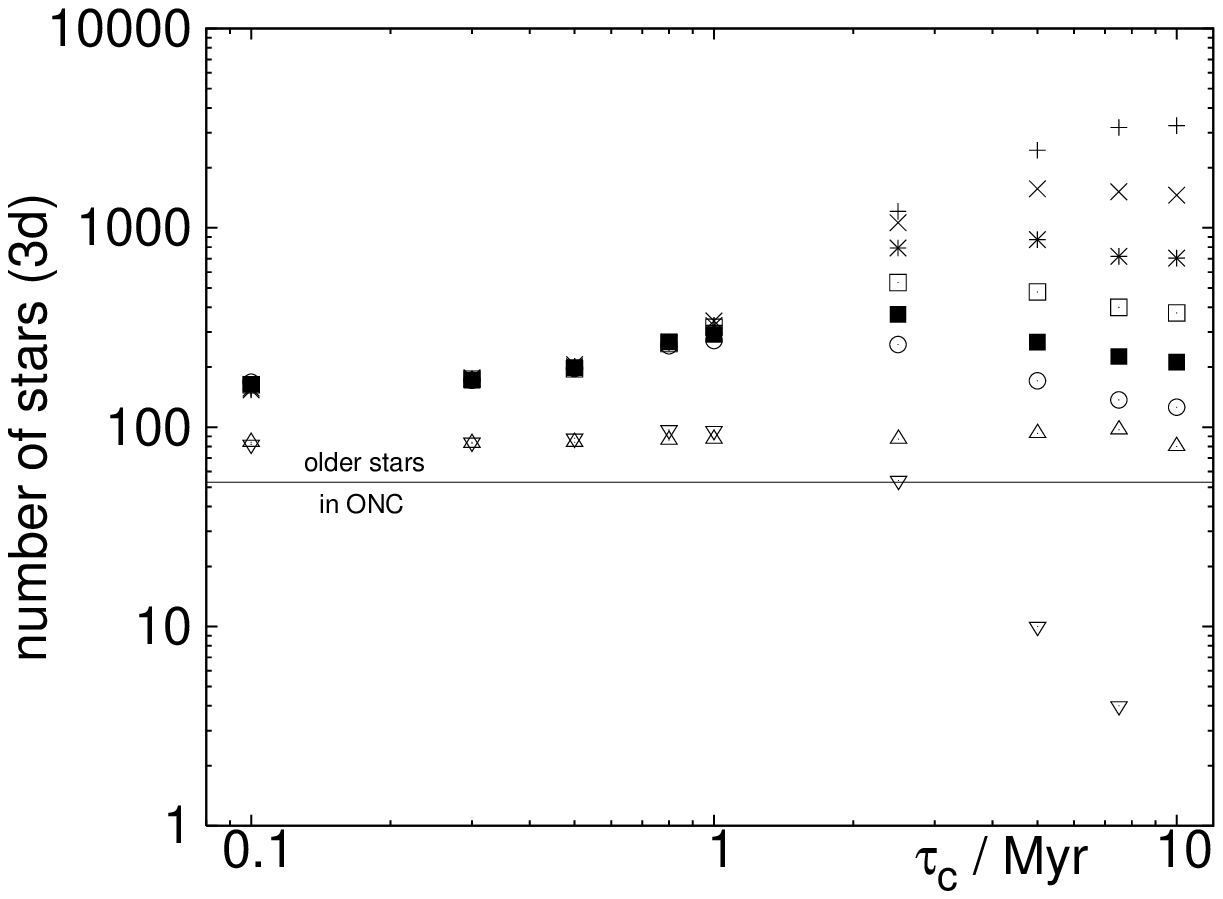}
  \caption{Number of stars within 2.5~pc of the centre after 
    the collapse time-scale, $\tau_\mathrm{c}$, has elapsed and the
    collapse stops.
    \newline
    {\it Top}: Number of stars within 2.5~pc of the centre of the potential
    {\it and} gravitationally bound.
    {\it Bottom}: Number of stars within 2.5~pc of the centre of the 
    potential.
    {\it Symbols}: Models with a one-dimensional  
    velocity dispersion, $\sigma$, 
    of the background field in presence of the collapsing cloud:
    $\sigma = $ 
    0.5~km~s$^{-1}$ (\protect\includegraphics[width=2ex]{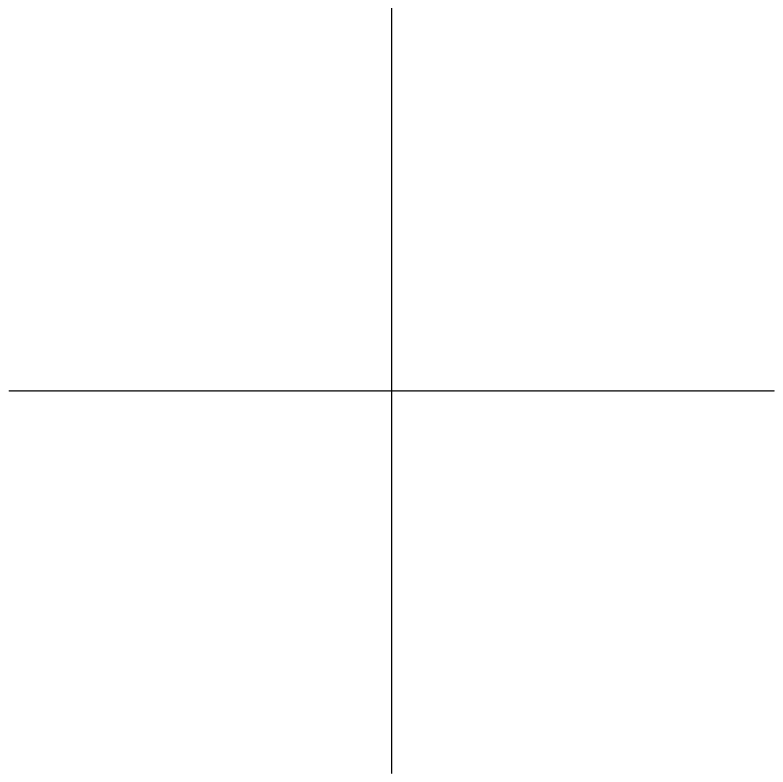}),
    1.0~km~s$^{-1}$ (\protect\includegraphics[width=2ex]{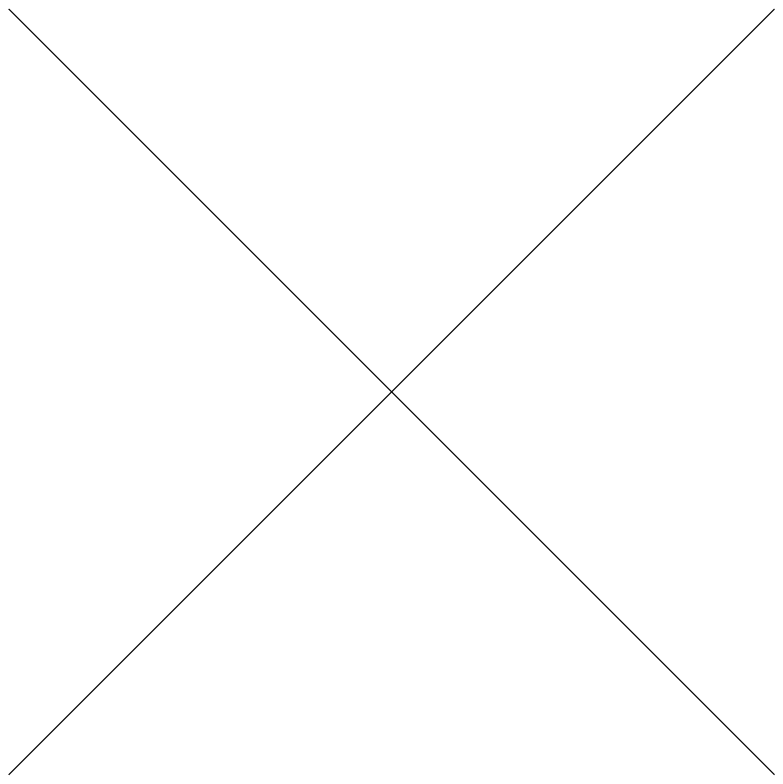}),
    1.5~km~s$^{-1}$ (\protect\includegraphics[width=2ex]{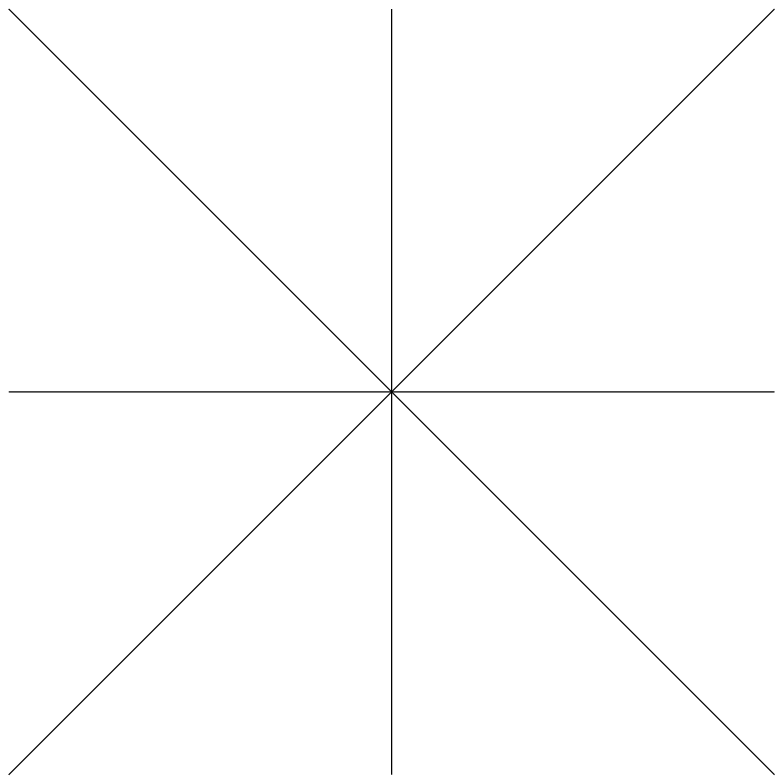}),
    2.0~km~s$^{-1}$ (\protect\includegraphics[width=2ex]{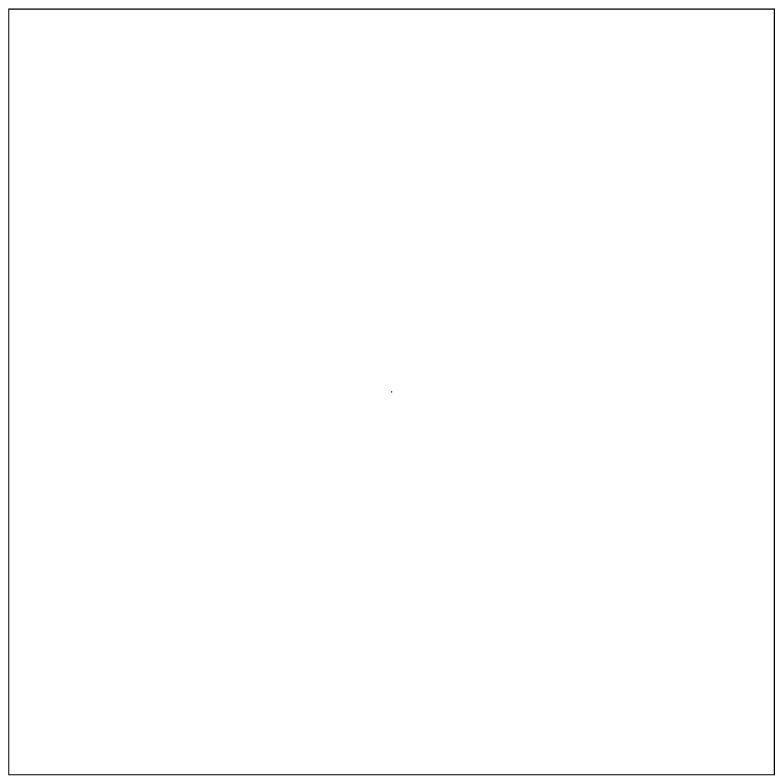}),
    2.5~km~s$^{-1}$ (\protect\includegraphics[width=2ex]{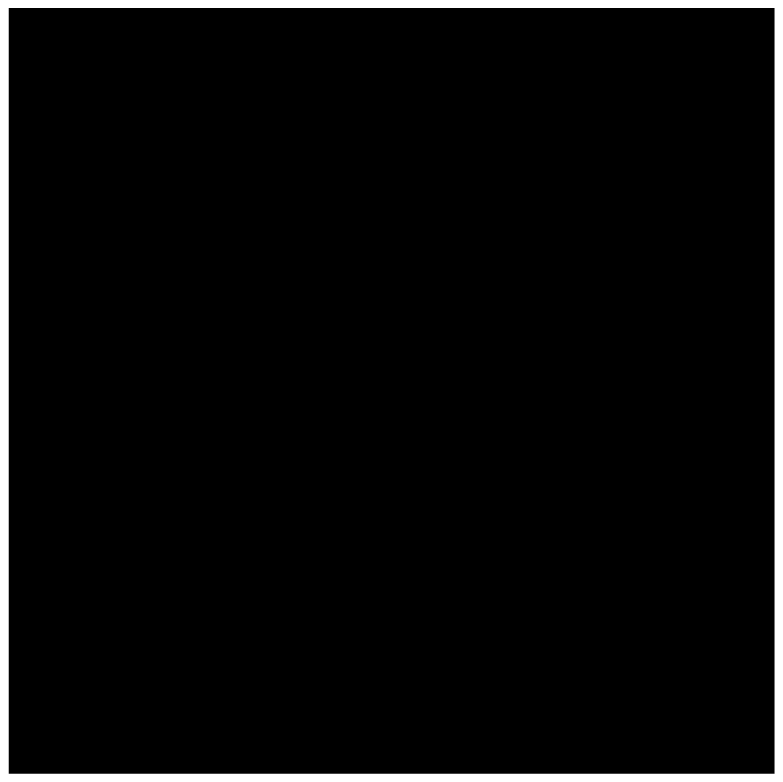}),
    3.0~km~s$^{-1}$ (\protect\includegraphics[width=2ex]{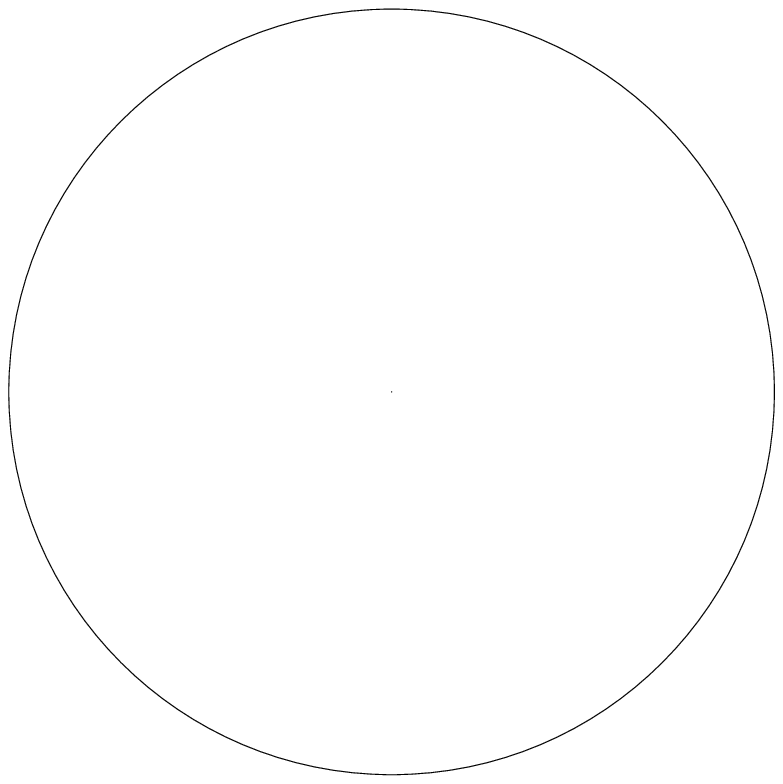}).
    The open triangles denote the number of stars within 2.5~pc radius
    of the centre of the background sphere with time if {\it no} 
    cloud potential is present for two different field-star velocity
    dispersions: 
    $\sigma = $
    0.5~km~s$^{-1}$ (\protect\includegraphics[width=2ex]{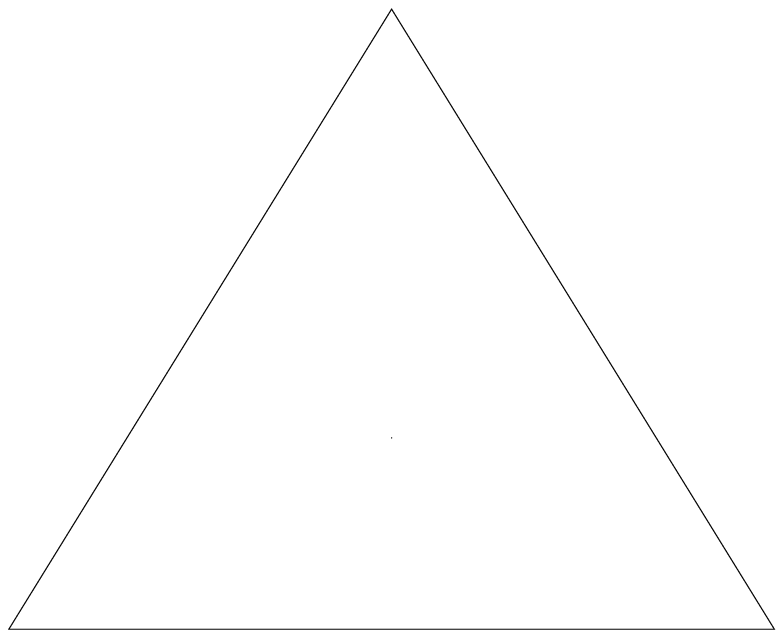})
    and
    3.0~km~s$^{-1}$ (\protect\includegraphics[width=2ex]{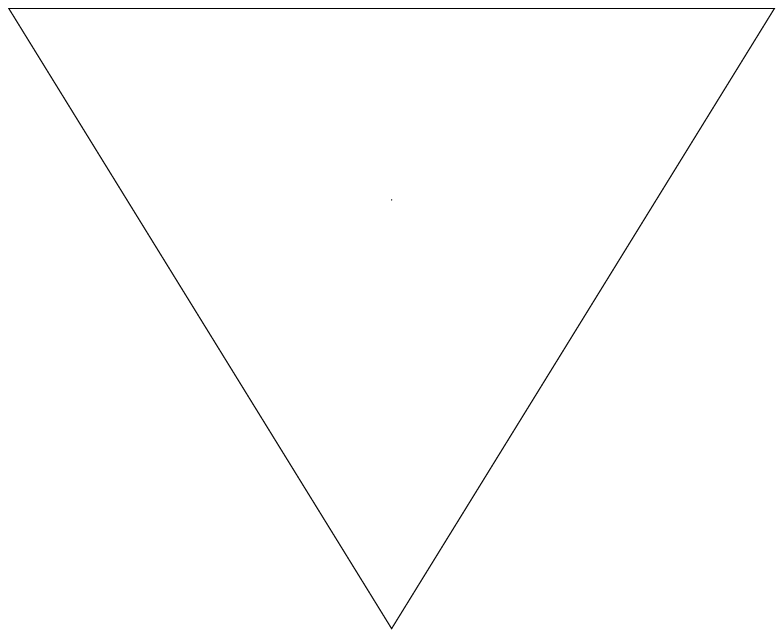}).
    The horizontal line marks the region of the
    53 expected older low-mass stars in the ONC.
    For details see text.
}
\label{captured_stars}
\end{figure} 

The number of captured stars in dependence of the collapse time-scale,
$\tau_\mathrm{c}$, can be seen in Fig. \ref{captured_stars}.
The initial stellar density is 2.44~stars/pc$^{3}$ implying 
a total number of stars placed initially within a 2.5~pc radius around
the centre of the potential of 160 
(Section \ref{underlying_field_population}). 
A decreasing collapse time means that the time-dependent
period becomes shorter and the potential behaves increasingly as
an instantaneously switched on  
potential. Therefore only those stars being initially within
the 2.5~pc sphere can be captured for short $\tau_\mathrm{c}$. Out of these
candidates only that fraction of stars is captured having a velocity
less than a certain limit. Thus the models with the shortest 
collapse time of 0.1~Myr converge against the initial star number of 157
with decreasing field-star velocity dispersion 
(Fig. \ref{captured_stars}, top).
With increasing collapse-time more stars are able to move inwards 
loosing energy in the time-dependent potential and the number of
captured stars increases. In the case of the highest field-star 
velocity dispersion
the number of captured stars turns over at large collapse times because
the potential is very low initially for a longer period. The diffusion of the 
background sphere then dominates over orbit focussing by the increasing
potential. 

To compare the number of captured stars calculated in these simulations
with the number expected it has to be taken into account that 
the test particles forming the background sphere 
represent a realistic stellar population having a mass spectrum.
Given a canonical IMF 
13.5 per cent of  all stars lie in 
the mass regime 0.4--1.0~M$_\odot$.
For the preferential value of the collapse time of 5~Myr the number of
captured stars is 42 (510) if the velocity dispersion of the
background sphere is 3.0 (0.5) km~s$^{-1}$, resulting in 6 (69)
captured low-mass stars.

Given its age of about 1~Myr the ONC is already dynamically evolved
and has expelled its gas almost completely.
Its virial mass of about 4500~M$_\odot$ is more than twice larger than
the stellar mass. Being nearly virialized at its formation the ONC
has already started to expand implying that its concentration
must have been larger initially and its describing Plummer parameter
$b_\mathrm{0}$ must have been smaller than 0.3~pc, 
when cloud collapse has stopped. 
This has been confirmed by numerical simulations \citep{kroupa2001b}.
Thus the potential would have been deeper 
than assumed in the present calculation, 
and the number of captured stars would
be higher than determined here.

Therefore, the existence of older stars in the 
ONC does not necessarily imply that star formation is prolonged -- 
dynamical capture can explain the presence of older stars in the 
ONC and in young star clusters in general. 

From the dynamical point of view these gravitationally bound stars 
are true cluster members. It is not a question of measurement accuracy
that these stars are identified to be cluster members. So even GAIA
can not distinguish between the older stars captured by cloud collapse
and the stars formed in the young star cluster, although on average
the captured stars would have a flatter density profile than the stars
formed in the cluster \citep{fellhauer2006a}. 

\revised{As the membership probability of the sample of low-mass
stars in the ONC, selected by \citet{palla2005a}, is greater than 90 per
cent but not 100 per cent,
based on proper motion and radial velocity studies, it is possible that
these older stars (or some of them)
are part of the fore- and background contamination.
But independently of the true origin of these  older stars,
stellar capture during pre-cluster cloud collapse must be a true
physical process and older stars among newly formed stars in a young star
cluster should exist.}
\section{Enhancement of stellar density}
In the previous section only those stars were
 considered which are gravitationally
bound. These constitute a fraction of all those field-stars which are
deflected from their initial orbit and are focussed towards the centre of
the cluster. In general, the density of the field-stars will deviate
from their constant initial density in two ways.
\subsection{Slowed-down dilution of an OB~association}
In absence of a collapsing pre-cloud cluster core the OB~association
will disperse due to its internal velocity dispersion. In presence of
such a collapsing cloud the stars are increasingly attracted towards
the centre of the newly formed cluster. Therefore the field-star density 
near the cluster centre will be kept higher during the cluster formation
than in absence of the cluster potential. In the case of a collapse-time
of 5~Myr and a field-star velocity dispersion of 3~km~s$^{-1}$ 
the number of stars within 2.5~pc of the centre of the background
sphere is 108 (with cloud collapse) and 10 (without cloud collapse)
when the collapse stops.
This means that the field-star density will be kept higher 
by a factor of 10 
due to the attracting potential by the collapsing cloud
(Fig. \ref{captured_stars}, bottom).

\subsection{Underestimation of  background subtraction} 
\begin{figure}
  \includegraphics[width=\figurewidth]{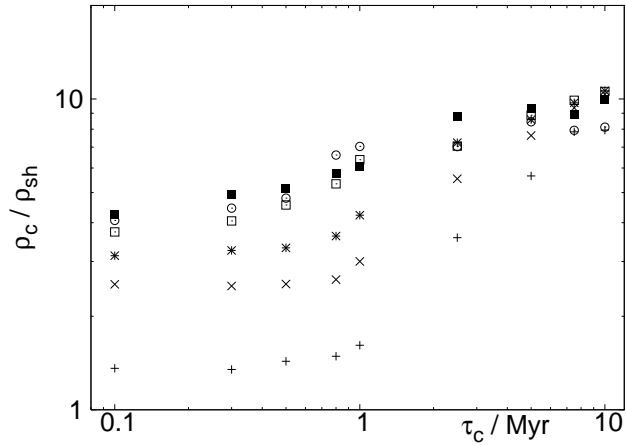}
  \caption{Ratio of the field-star density in the cluster, 
    $\rho_\mathrm{c}$, and in the
    vicinity of the cluster, $\rho_\mathrm{sh}$, 
    after the collapse has stopped.
    {\it Density in the cluster}: Number of stars within 2.5~pc radius 
    of the cluster centre divided by the volume of the sphere.
    {\it Density in the vicinity}: Number of stars in the shell
    between 10~ and 7.5~pc divided by the volume of the shell. 
    {\it Symbols}: Models with a one-dimensional  
    velocity dispersion, $\sigma$, 
    of the background field in presence of the collapsing cloud: $\sigma = $
    0.5~km~s$^{-1}$ (\protect\includegraphics[width=2ex]{./pointtype1.eps}),
    1.0~km~s$^{-1}$ (\protect\includegraphics[width=2ex]{./pointtype2.eps}),
    1.5~km~s$^{-1}$ (\protect\includegraphics[width=2ex]{./pointtype3.eps}),
    2.0~km~s$^{-1}$ (\protect\includegraphics[width=2ex]{./pointtype4.eps}),
    2.5~km~s$^{-1}$ (\protect\includegraphics[width=2ex]{./pointtype5.eps}),
    3.0~km~s$^{-1}$ (\protect\includegraphics[width=2ex]{./pointtype6.eps}).
}
\label{background_substraction}
\end{figure}

When the pre-cluster cloud becomes increasingly compact
then the radial dependence of its potential increases.  
The density of the field-stars is expected to show a similar 
dependence, getting higher towards to the cluster centre. The corresponding
ratios of the field-star densities in the newly formed cluster
and the cluster vicinity are plotted in 
Fig.~\ref{background_substraction}. After the collapse has stopped
the cluster density, $\rho_\mathrm{c}$, 
is calculated by the number of field stars within 2.5~pc
radius of the cluster centre. The density of the 
field-stars, $\rho_\mathrm{sh}$, 
in the cluster
vicinity is calculated by the number of stars within the shell limited
by the radii of 10.0 and 7.5~pc. For a collapse time of 5~Myr and a
field-star velocity dispersion of 3~km~s$^{-1}$ the density of the 
field-star in the new cluster is almost 10~times higher 
than in the vicinity of the cluster. This has to be taken into account
when surveys of star clusters are corrected by the fore- and 
background contamination. 

In general three tendencies can be seen. 
i )The density contrast  increases with an increasing collapse-time.
ii) For a constant collapse time the density contrast  
    increases with an increasing initial 
    velocity dispersion of the background population.
iii) With increasing collapse-time the final density ratios for different
initial velocity dispersions are less different.

The behaviour of the data in Fig. \ref{background_substraction}
can be understood by noting the following:
If the full potential would be switched on 
instantaneously then the density contrast
would be unity. With an increasing collapse-time 
more stars can move towards the cluster centre and the density contrast
increases. Within  a constant collapse-time more stars can move towards the
cluster centre if the initial velocity dispersion is higher.
\section{Discussion}
\revised{
\label{discussion}
As the region, where the ONC lies in, is a location of high
star formation, the initial conditions prior to pre-cluster cloud
collapse, i.e. the stellar density and distribution, 
are difficult to estimate. If the older ONC stars have not formed
in the ONC as a result of prolonged star formation 
then where could they have formed?

Given the age of the association Orion OB1c 
surrounding the ONC of about 4.6~Myr
\citep{brown1994a} the observed older stars in the ONC, having an age 
greater than 
10~Myr, can not have formed in the parent cluster of this association.

The association Orion OB1a has an age of approximately 11~Myr. The projected
distance from its centre to the ONC is of about 5 degrees, i.e. 35~pc
given a mean distance of 400~pc of both associations. The line-of-sight
distance between Orion OB1a and the ONC is approximately 117~pc
\citep{brown1999a}. Then the probably captured 
low-mass stars, if formed in the association
Orion OB1a, must have had a spatial velocity of 12~km~s$^{-1}$ relative
to the centre of the ONC for them to have drifted to the current distance of
the ONC. If the time-scale of the pre-cloud core
collapse is taken into account the velocity must be even higher.
Stars from such a population would not have been captured.
Additionally, \citet{brown1999a}
listed 61 OB stars for the Orion OB1a association. Given a canonical IMF
the parent cluster should have contained about
3150~M$_\odot$ in stars and therefore 1100 stars between 0.4 and
1.0~M$_\odot$. If all these low-mass stars are completely 
radially dispersed during the expansion of the OB-association
the stellar flux at  the  ONC would have been $6\times
10^{-3}$~stars/pc$^2$. Given a radius of 2.5~pc of the ONC then 0.5
stars with masses between 0.4 and 1.0~M$_\odot$ have passed through
the ONC, too few compared to the estimated number of older low-mass
stars of 53. It seems to be unlikely that these older stars have formed
in Orion OB1a, although its age would be consistent with this.  

The nearer association Orion OB 1b has an age of 1.7~Myr \citep{brown1994a}.
This is contrary  to the ages determined by \citet{blaauw1999a} (7~Myr)
and by \citet{warren1978a} (5.1~Myr). Nevertheless, it can be concluded that 
Orion OB 1b might be too young to be the origin of the older
stars in the ONC. 
 
The compact ONC has 3500 stars within a radius of 2.5 pc
around $\Theta^{1}$ C \citep{hillenbrand1997a}, 
giving a number density of 528~stars~pc$^{-3}$. 
As the current ONC has an age of approximately 1~Myr it is already
dynamically evolved. Using full $N$-body simulations
including gas expulsion \cite*{kroupa2001b} have shown that initially 
the ONC may have contained approximately 10$^4$ stars and brown dwarfs
to match its current state. This means that the 
ONC has already lost 65 per cent
of stars. To get 20000 stars within a sphere of 25~pc in diameter
2 ONC-type cluster are required. This huge amount of past star formation
should still be observable in terms of an association. 
At least the $2\times 39 = 78$ O-stars,
given a canonical IMF,  should still be visible. Given the live time
of a 20~M$_\odot$ star of approximately 9.9~Myr \citep{schaerer1993a},
then $2\times 10 = 20$ supernovae should have occurred, where 
10~$m\ge20\;\mathrm{M}_\odot$ stars occur in a population of 
39~$m\ge\;\mathrm{M}_\odot$ stars.
Reducing the radius
of the background sphere to 9.9~pc then only 10000~stars or one
ONC-type star cluster would be 
required to keep the initial stellar density of 2.4~stars/pc$^{3}$.
Using a canonical IMF the mass of a star cluster must be less than 
160~M$_\odot$ to contain less than 1~O star. Nearly 20 such low-mass
clusters are required to produce $2\times10^4$~stars in total. 
If the  background
sphere had only a radius of 9.9~pc then only 10 such low-mass clusters 
are needed to form the initial background density.
If these
low-mass clusters formed more than 10~Myr ago, then they are not
expected to be still visible, as such small star clusters
disperse rapidly. Also, many of these older stars should be mixed
up with the ONC. 

However, our results do not exclude the possibility that the ONC formed
over about 3 Myr, i.e. a few free-fall times \citep{tan2006a}. The
important time scale for the effects described here is the time scale of
potential formation and not the time scale of star formation.


Possibly, the most likely source of the older low-mass star population
might be a large number of small groups producing 
low-mass stars as are also observed today but only 10~Myr 
earlier \citep{megeath2005a}. For example, \citet{kroupa2003a}
have shown that small Taurus-Auriga-type groups disperse on a 
time scale of a few Myr.

In our initial conditions it is assumed that the ONC and the
background sphere have the same velocity centroid. If 
the relative velocity between the background sphere and the collapsing
pre-cluster cloud core increases then fewer stars are expected to be
captured. But the pre-cluster cloud core of the ONC as well as
the earlier star clusters or groups, which formed the background stars, 
may have formed within the same molecular cloud.
So a relative velocity between the different velocity centroids
may have been small.   

The increasing potential of the collapsing pre-cluster cloud core should not
only act on already formed stars but also on protostars. Applying the
results of the enhancement of the background density on the
stellar population in the molecular cloud Orion A 
the mean motion of the young stars and the protostars
should be directed towards the
centre of the ONC, being more pronounced in the region of the
cloud nearest to the ONC.
}
\section{Conclusions}
We have shown that the existence of older stars in the ONC must not
necessarily imply that star formation was prolonged in the ONC.
The time-dependent potential of a collapsing cloud can capture
older stars of the underlying OB~association. These stars
exhibit similar  kinematical properties leading to their
identification as cluster members. The number of captured
stars is in agreement with the number of observed older low-mass
stars for reasonable assumptions about the pre-existing field-star
population. Nevertheless, some open issues concerning the origin
of the field-star population remain.

Additionally, the increasing potential of the collapsing cloud 
leads to an enhancement of the local stellar background density
causing larger fore- and background contamination. 

The number of captured stars and the amount  of fore- and background
enhancement significantly depend on the deepness of the final
potential when the cloud collapse stops. Therefore these 
effects should be more distinctive for more compact and more massive
star clusters \citep{fellhauer2006a}.

Moreover, the number of capture stars and the number of focussed but not
captured stars
depend on the surrounding density of the field stars. 
Therefore the number of older stars  found in
young star clusters that are part of a greater star-formation region
and/ or are embedded in an OB association should be higher than in 
young star clusters formed in isolation.

We conclude that the apparent long formation time of young star
clusters may be brought into agreement with the recent notion that 
the formation of clusters is a highly dynamic and violent process by taken
into account that the short formation process leads to stellar capture 
from an underlying older field population. This process depends on the
time-scale of cluster potential formation, the velocity dispersion and density
of the field population.

\vspace{1cm}
This work was supported by the GRK-787 Bochum-Bonn 
\emph{Galaxy groups as laboratories for baryonic and dark matter}.
JP-A thanks especially Ralf J\"urgen Dettmar, spokesman
of the GRK-787. 
We also thank Thomas Preibisch for useful discussions concerning 
OB associations.
\bibliographystyle{mn2e}
\bibliography{n-body,ONC,star-formation,imf,milkeyway,star-cluster,stellar-evolution}

\end{document}